# Representing Born effective charges with equivariant graph convolutional neural networks

Alex Kutana[1], Koji Shimizu[2], Satoshi Watanabe[3] and Ryoji Asahi[1] ✉

[1]Nagoya University, Furo-cho, Chikusa-ku, Nagoya, Japan
[2]National Institute of Advanced Industrial Science and Technology (AIST), Tsukuba, Japan
[3]The University of Tokyo, Hongo, Tokyo, Japan

✉ **E-mail:** asahi.ryoji.d9@f.mail.nagoya-u.ac.jp

## Abstract
Graph convolutional neural networks have been instrumental in machine learning of material properties. When representing tensorial properties, weights and descriptors of a physics-informed network must obey certain transformation rules to ensure the independence of the property on the choice of the reference frame. Here we explicitly encode such properties using an equivariant graph convolutional neural network. The network respects rotational symmetries of the crystal throughout by using equivariant weights and descriptors and provides a tensorial output of the target value. Applications to tensors of atomic Born effective charges in diverse materials including perovskite oxides, $Li_3PO_4$, and $ZrO_2$, are demonstrated, and good performance and generalization ability is obtained.

## 1. Introduction

Since their introduction in materials science and chemistry, graph convolutional neural networks (GCNNs) have shown remarkable results[1]. GCNNs are being used to represent various molecular and materials properties[1–3], with typical tasks of predicting scalar values for electronic band gaps, ionization potentials/electron affinities, formation energies, as well as performing classification. In particular, in interatomic potentials and force fields applications, GCNNs presently achieve state of the art performance[4–6]. While most applications have been limited to scalar target values, many of the fundamental properties of materials are represented by tensors. When constructing a neural network with tensorial outputs, it is essential to incorporate relevant physics-informed[7] constraints and inductive biases, thus greatly improving the model performance in comparison with data augmentation approaches[8,9].

Tensors must obey certain transformation rules to ensure the independence of physical properties on the choice of the reference frame[10]. In order for the network features and outputs to follow these rules, it is necessary to encode their equivariance explicitly into the network architecture. Stemming from the original idea of steerable filters for vision and image processing[11], feature equivariance ensures their proper transformations in convolutional neural networks[12,8,13–15]. In physics and materials science, equivariance with respect to translations and rotations in three-dimensional Euclidean space is most essential. Utilization of Euclidean equivariant networks[9] has greatly increases the accuracy of interatomic potentials[5,6], and benefited other tasks, including prediction of optical, phonon, and neutron scattering spectra[16–18], density functional Hamiltonians[19,20], ab initio wavefunctions[21], reaction activation barriers[22], and spin dynamics[23].

At the same time, direct, end-to-end applications of machine learning to tensorial atomic quantities have been scarce. Recent developments include predictions of NMR chemical shift[24] and elasticity[25,26] tensors. Our earlier efforts utilized a symmetry-restricted network to predict part of the tensor of atomic Born effective charges[27]. The network was designed for a special case of the external field directed along one of the Cartesian axes, and accommodating the general case would have been substantially more complex. Here, we report Equivar, a simple equivariant GCNN (EGCNN) for predicting atomic tensorial quantities, which observes the symmetries of the physical three-dimensional Euclidean space, and operates with geometric quantities throughout. It utilizes local message passing based on the geometric input data. As only equivariant operations are performed on the input geometric data, all constraints imposed by the point group symmetry are automatically satisfied by the outputs. We demonstrate the application of Equivar to end-to-end machine learning of the target property of full tensors of atomic Born effective charges, achieving good performance. Due to equivariance constraints, high accuracy is achieved, while keeping the number of parameters small.

## 2. Network architecture

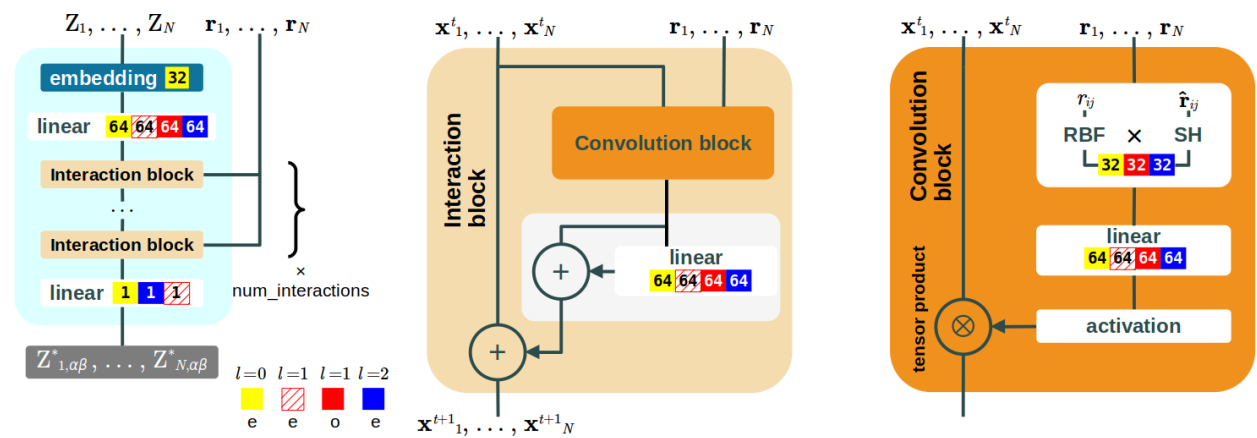

**Fig. 1 | Architecture of the Equivar EGCNN.** The inputs are atomic numbers and positions, and outputs are full tensors of atomic Born effective charges. $l=0$, $l=1$, and $l=2$ equivariant features are colored yellow, red, and blue, respectively; "e" and "o" stand for even and odd parity. Numbers indicate feature multiplicities. The values shown correspond to those used in "base model 1".

The atomic structure is represented by a graph[28], with atoms mapped to graph nodes and neighbor connections to edges, each equipped with attributes/features. Node features are updated via message passing from the neighbor nodes in the interaction block.

The basic architecture of Equivar is shown in Fig. 1. The input attributes are scalars (atomic numbers $Z_i$, interatomic distances $r_{ij}$), and vectors (bond directions $\hat{\mathbf{r}}_{ij}$), while the output features are the node values of tensors of atomic Born effective charges $Z^*_{i,\alpha\beta}$. Node embedding is a lookup dictionary with the mapping of the atomic numbers into an array of learnable scalar weights. The embedding dimension is a hyperparameter. Edge attributes represent the length and direction of the vector connecting neighbors. Edges are indexed by the bond vector $\mathbf{r}_{ab}$ connecting the central atom *a* with neighbor atom *b*. One-hot edge encodings $\delta(\mathbf{r}-\mathbf{r}_{ab})$ are projected onto a fixed basis of products of radial and angular functions, yielding the mapping $\mathbb{R}^3 \to \mathbb{R} \times \mathbb{R}^{2l+1}$ for a radial basis function $R_n(r)$ and a set of angular basis functions of real spherical harmonics $Y_{l,m}(\theta,\varphi)$. A similar expansion is employed to obtain the rotational power spectrum and smooth overlap of atomic positions (SOAP) similarity kernel[29]. Uniformly shifted Gaussians $R_n(r)=\exp(-\gamma(r-\mu_n)^2)$ [30] were used as radial basis functions[31], and a 3 Å radial cutoff was used for neighbor lists. The set of spherical harmonics $Y_{l,m}$ of degree *l*, $|m|\leq l$, produces equivariant geometric edge features enabling the "information flow"[13] within the channels of a geometric object upon rotation. The projections are passed through linear and activation blocks, producing a trainable tensor product convolution kernel. To provide a nonlinearity, a shifted softplus activation function[30], $\log(1+e^x)-\log(2)$, is applied to the scalar edge features.

 After the embedding, an equivariant linear layer introduces geometric node features, initialized to invariant (zero) values. The geometric (equivariant) features are irreducible representations of *O*(3), a three-dimensional orthogonal group. Unlike the invariant (scalar) features, the equivariant features are "space-aware" and transform with rotations according to certain rules. Encoding these rules into the weights of the neural network increases its accuracy significantly[7], as the target quantities share same transformation properties as tensors. The features are stored sequentially in conventional arrays of floats, and typed according to how they transform under *O*(3)[9]. The equivariant linear layer acts by taking linear combinations of irreducible representations with the same rotational order *l* and parity *p*, and storing them in each of the matching outputs. The scalar coefficients of the linear combinations are learnable parameters. The equivariant node features are then updated via message passing in a sequence of interaction blocks, with the number of interactions being a hyperparameter. Within the interaction block, the message passed from a node to its neighbor is a tensor product of the node features with the features of the edge connecting the nodes, and the messages are aggregated by target nodes in a ResNet-style update[32]. The tensor product is calculated using Clebsch-Gordan coefficients[15]. For convenience, identical descriptor types and sizes for node and edge are used in the interaction block.

After the interactions, the final linear layer yields a decomposition of a tensor into irreducible representations of the *O*(3) group, namely 0e+2e+1e, which has 9 components that are converted to Cartesian form. For equivariant operations with geometric features, the primitives implemented in `e3nn`[9] are used.

## 3. Datasets

The equivariant model was trained with full tensors of Born effective charges, calculated from first principles using density functional perturbation theory (DFPT)[33], as implemented in the VASP package[34]. PBEsol functional[35] was used for perovskites, and PBE functional[36] for $Li_3PO_4$ and $ZrO_2$. Ion-electron interactions were represented by all-electron projector augmented wave potentials[37], and a plane wave basis was used with cutoff energy of 500 eV ($Li_3PO_4$) and 520 eV (perovskites and $ZrO_2$). Tensors of atomic Born effective charges are fundamental quantities determining the long-range long-wavelength part of the force constants, LO-TO gamma point optical phonon splitting, and static dielectric response[33,38]. They are defined as the linear part of the change in polarization due to gamma point ion displacement, and given by the mixed second derivatives of the total energy $E$ with respect to atomic positions and electric field[33,39–42]:

$$Z^*_{i,\alpha\beta} = -\frac{1}{|e|}\frac{\partial^2 E}{\partial \mathcal{E}_\alpha \partial u_{i,\beta}} \tag{1}$$

Here, $Z^*_{i,\alpha\beta}$ is the tensor of the Born effective charge of atom $i$, $\mathbf{u}_i=\mathbf{u}_i(q=0)$ is the gamma point displacement of atom $i$, $e$ is elementary charge, and $\mathcal{E}$ is the electric field. A definition with swapped Cartesian indices $\alpha$ and $\beta$ has also been used[43–45]. In the DFPT approach, the ground state and first derivatives of the wavefunction are calculated to obtain the Born effective charges. After training with DFPT results, the equivariant end-to-end ML model quickly predicts the Born charges tensors from the structural input.

We use three datasets representing different systems of interest for training. The datasets were generated in-house by performing substitutions or creating vacancy defects in pristine bulk structures obtained from the Materials Project[46] database. Structure ids of the structures used are given in the Supporting Information. The first dataset contains substituted perovskite oxides, relevant to electric energy storage applications[47]. Substitutions are a primary method for boosting the dielectric permittivity of transition metal oxides[48–50], leading to greater energy storage capacity. Model validation with this dataset also demonstrates its capability for a wide range of chemical elements. The dataset was generated via cation substitutions in mineral perovskite, Pnma $CaTiO_3$. $Ca^{2+}$, $Sr^{2+}$, $Ba^{2+}$, and $Pb^{2+}$ isovalent substitutions were performed on the alkaline earth metal site, and $Ti^{4+}$, $Zr^{4+}$, and $Hf^{4+}$ on the transition metal site[49,50] to generate a dataset of 1,224 materials. Born charges of optimized structures were used for training. The second dataset contains various structures of the Li-ion battery material, $Li_3PO_4$, one of the most widely used solid electrolytes[51–53]. The dataset of MD snapshots including defects may test sensitivity of local configurations. This dataset was used to elucidate the Li ion conduction behaviors[27,53]. Pristine $Li_3PO_4$ ($Li_{12}P_4O_{16}$ in the adopted supercell), as well as systems with Li and $Li_2O$ vacancy defects ($Li_{11}P_4O_{16}$ and $Li_{22}P_8O_{31}$, respectively)[27] were used. The pristine subset consists of snapshots from NVT-ensemble ab initio molecular dynamics (AIMD) simulations at 300 and 2,000 K. A time step of 1 fs was used. The Li vacancy structure set contains images from nudged elastic band (NEB) calculations, whereas $Li_2O$ vacancy structures are snapshots of AIMD at 2,000 K. All frames from the ab initio MD calculations were used. The total number of systems is 17,991; additionally, a dataset with 1,870 larger structures ($Li_{46}P_{16}O_{63}$) with a $Li_2O$ vacancy was used for testing. The third dataset has structures of zirconia ($ZrO_2$), a high permittivity material[54], with applications in microelectronics, energy storage, and as structural ceramics[55]. The dataset with three different crystal structures — cubic, tetragonal, and monoclinic — contains the subtle effects of these long-range orders. It is planned to use this dataset to analyze the mechanism of plastic deformation enhancement under electric field application[56]. Materials Project structure ids of the structures used are given in the Supporting Information. The dataset consists of 10,103 NVT-ensemble empirical potential simulation[57] snapshots of cubic, tetragonal, and monoclinic zirconia ($Zr_{16}O_{32}$ in the supercell for the pristine models) at 1,300, 1,500, 1,700, and 1,900 K, with isotropic lattice constant changes (-2%, -1%, 0%, +1%, +2%), and with some of the systems hosting an oxygen vacancy. For the oxygen vacancy structures, +2 charge state was considered. The snapshots were taken every 1000 fs.

## 4. Results

We benchmarked the model while tuning the hyperparameters using a small dataset of 224 perovskites. The model was trained for 250 epochs using $L^1$-norm loss function and AdamW optimizer[58] with an initial learning rate of 0.005, and best validation mean absolute error (MAE) was recorded. The results are given in Supplementary Table 1. The model with ~131k trainable parameters (6 interaction layers and feature size of 32 in the interaction block) has the MAE of 0.0117 for the small set validation, compared to the cross-validation MAE of 0.0088 for the full set of 1,224 perovskites. The model demonstrates a good generalization ability even with small training data, and stable performance with respect to hyperparameter variation. The performance can be further improved by increasing the model size and the number of interactions. Here, we use a 510k-parameter model, 'base model 1' (BM1), with 6 interaction layers and feature size of 64 in the interaction block, which provides a good tradeoff among the accuracy, speed, and size, and compares with the smaller/faster 131k-parameter model (BM2). The two models were trained for 500 epochs with a combined dataset of three systems types (perovskites, $Li_3PO_4$, and $ZrO_2$), as well as several subsets, using a 0.8/0.2 train/validation split. The training history with validation loss is shown in Fig. 2. BM1 shows consistently better performance, due to larger size of the descriptor in the interaction block. Further increase in the descriptor size leads to even better performance, but is not pursued due to memory constraints. Another important hyperparameter is the number of interactions. Iterative interactions allow the nodes to exchange messages beyond the nearest neighbor cutoff, which allows capturing long-range interactions. In case of 3 Å cutoff radius and 6 interactions, the effective cutoff radius for the total receptive field of each node is 18 Å, sufficient to represent most long-range interactions. Here, 6 interactions were used, although the accuracy kept improving even at 12 interactions, the largest value tested.

The scalings of the training and inference times with the dataset size are shown in Supplementary Fig. 1 and 2, respectively. Calculations were performed on a single NVIDIA RTX A6000 GPU. Training and inference times of Equivar are found to scale approximately linearly with both the number of atoms and number of model parameters. The training times are 0.493 ms/atom/epoch for BM1, and 0.156 ms/atom/epoch for BM2, whereas the inference times are 0.340 ms/atom for BM1, and 0.117 ms/atom for BM2. These timings can be compared with the result of 1.05 ms/atom/epoch for SchNet on the Nvidia GTX 1080 GPU[59].

The model performance summary is given in Table I. The results for the models BM1 and BM2 trained with all three datasets are compared with those for "specialized" models, which were trained with only one dataset. It is seen that a specialized model can be trained to achieve a better performance with a particular dataset. The results of five-fold cross-validation for the tensor of Born effective charges of 1,224 perovskites are shown in Fig. 3. For cross-validation, the dataset was randomly split into five approximately equal-sized subsets. Each of the subsets was sequentially used for validation, while the remaining subsets for training the model, and validation results were recorded in each run. Cross-validation MAEs for diagonal and off-diagonal tensor components and individual atomic species are shown in Supplementary Fig. 3.

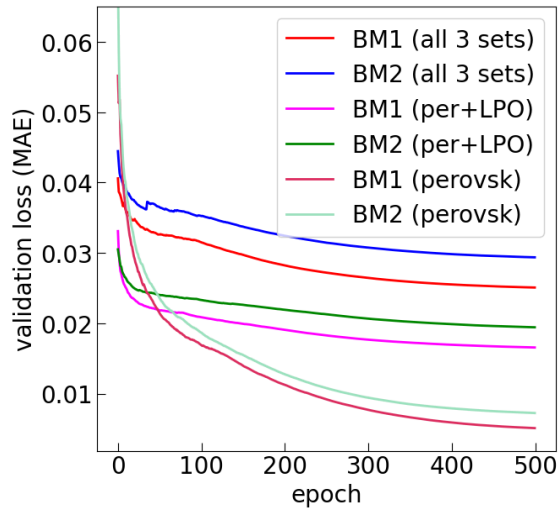

**Fig. 2 | Training history for EGCNN base models BM1 and BM2 for regression of tensors of atomic Born effective charges.** Models were trained for 500 epochs using combinations of dataset of three systems types - perovskites, $Li_3PO_4$, and $ZrO_2$. The larger BM1 model shows a consistently better performance.

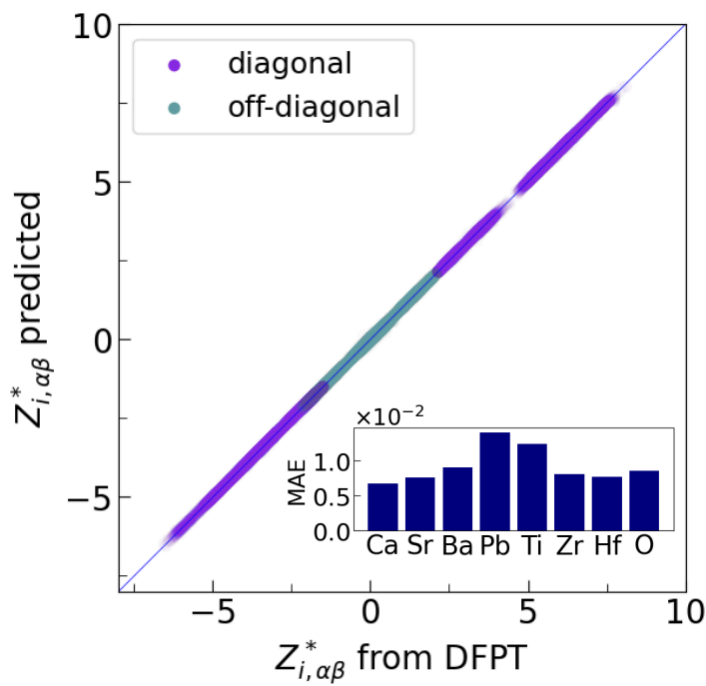

**Fig. 3 | Five-fold cross-validation results for the prediction of the tensor of Born effective charges of Pnma $ABO_3$ perovskites (A=Ca, Sr, Ba, Pb; B=Ti, Zr, Hf) by the equivariant GCNN model.** Diagonal and off-diagonal tensor components are shown. The inset shows the MAE for all components for individual atomic species.

The obtained diagonal and off-diagonal values for these elements are typical for the $ABO_3$ compounds[44,41]. Born effective charges of Ca, Sr, and Ba are rather similar, with average values for diagonal components of 2.5-2.7 and standard deviations of 0.15-0.17, whereas Pb has a larger average Born charge of 3.5 with a standard deviation of 0.26. Ti, Zr, and Hf have the average values of 6.5, 5.7, and 5.6, and standard deviations of 0.56, 0.41, and 0.42 respectively. The diagonal components for O range between -1.3 and -6.5, with the average of -2.9 and standard deviation of 0.91. Larger standard deviations in B-type cations and O as compared to A-type cations reflect the bimodal distribution of their Born charges, corresponding to two distinct oxidations states. Of note are the large variations of the off-diagonal components in O, ranging from -2.2 to 2.1, while in Ca the range is much smaller, from -0.26 to 0.34. The larger variability of the O values does not result in the larger error for the model prediction. Validation results for training with a $Li_3PO_4$ dataset are shown in Fig. 4. In contrast with the perovskites, the O Born charges in $Li_3PO_4$ show smaller values and variability, with the average being -1.53 and standard deviation 0.33 for the diagonal components. The average charge of Li is 1.06, corresponding to its nominal valence state, whereas that of P is 2.97. The prediction error is worse than for the perovskites dataset that only contains optimized structures without any defects. Thus, the larger prediction MAE in $Li_3PO_4$ is most likely due to the presence of anomalous values for a small number of structures in the dataset. The noise in the data could appear when atoms are displaced far from equilibrium in the high temperature MD.

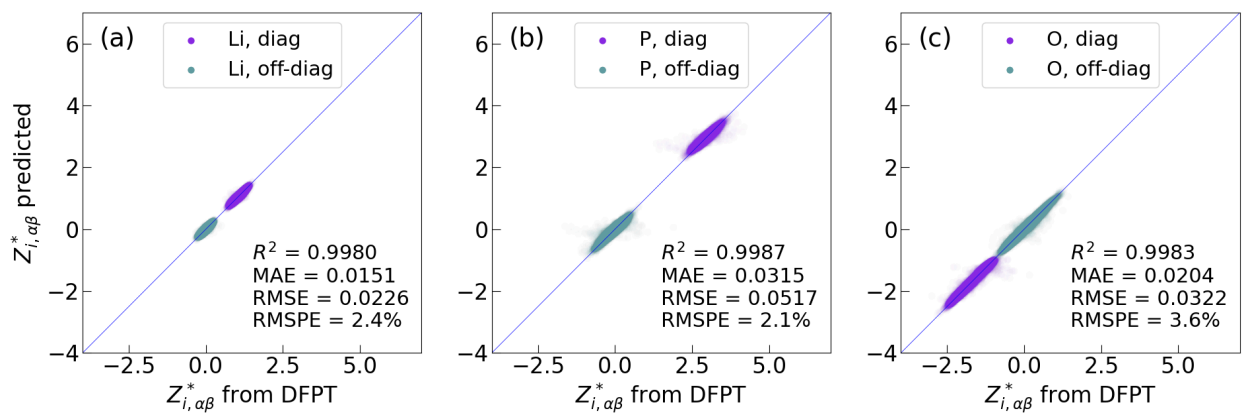

**Fig. 4 | Validation results for tensors of Born effective charges of $Li_3PO_4$.** Panels show results for different atomic species: **a** Li, **b** P, and **c** O. The structures used for model training are the snapshots of ab initio molecular dynamics of pristine $Li_3PO_4$ as well as structures with single Li or $Li_2O$ vacancies at 300 K and 2,000 K. RMSPE is given for diagonal elements.

Cross-validation results for the model trained with Born effective charges of $ZrO_2$ are shown in Fig. 5. In $ZrO_2$, the average O charge is -2.78, and std is 0.45, Zr average charge is 5.50 and std is 0.40. The error is larger in $ZrO_2$, which we surmised was due to the presence of a small number of anomalous data. We also trained an Equivar model using a cutoff radius of 5 Å and 3 interactions, to check whether a larger cutoff would better account for the nearest neighbor coordination and improve the performance. As seen from the validation plot in Supplementary Fig. 4, the model performance in $ZrO_2$ is similar to that of the original model, showing that the cutoff radius of 3 Å is adequate.

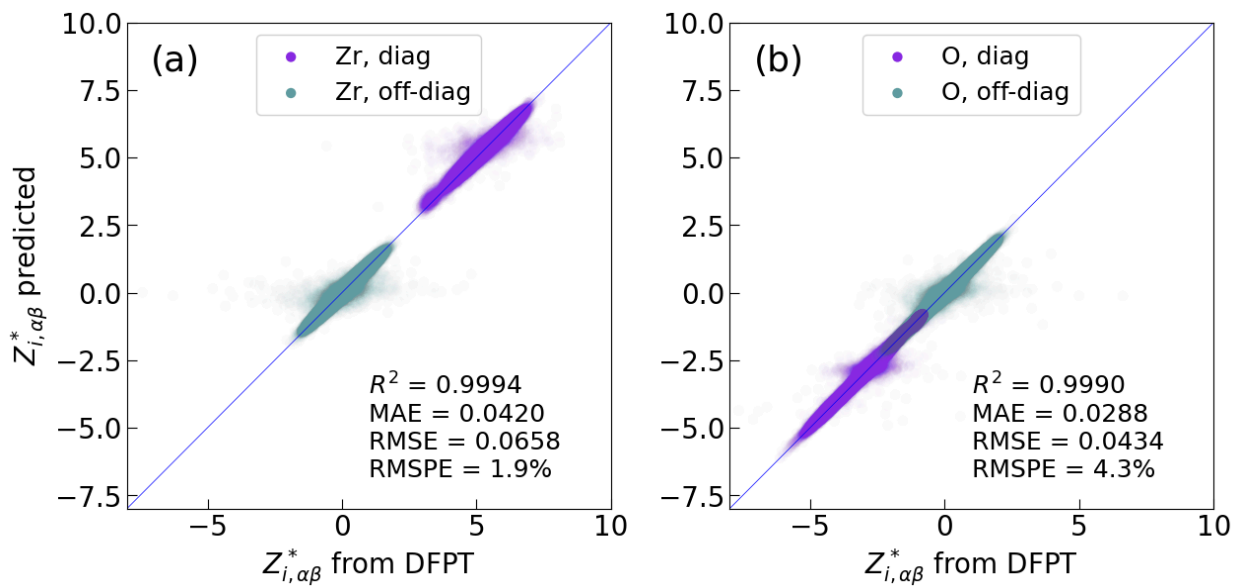

**Fig. 5 | Five-fold cross-validation results for the tensor of Born effective charges of ZrO$_2$.** The dataset consists of AIMD snapshots of cubic, tetragonal, and monoclinic zirconia, with some of the systems hosting an oxygen vacancy. Panels show results for different atomic species: **a** Zr, **b** O. RMSPE is given for diagonal elements.

## 5. Discussion

The Equivar model demonstrates good performance and generalization ability while retaining simplicity and small size. The model is comparable in size to SchNet[30], which has ~121k parameters (taking radial basis size of 32, same as used here, rather than 300 in the original SchNet). The larger size of Equivar is mainly due to its use of equivariant vector/tensor weights and features, which have larger dimensions than SchNet's scalars. Overall, while showing good performance, Equivar implements a rather minimalistic EGCNN for tensor representations, and further improvements are possible by employing more complex architectures. Recently, an equivariant model for regression of atomic Born effective charges based on λ-SOAP descriptors[60] has been reported[61]. The model archives a root mean square percentage error (RMSPE) of 3% for Born effective charges isolated water dimers. Equivar shows a better or similar accuracy for bulk systems, with RMSPE of 0.5%, 3%, and 4% for the diagonal elements of effective charge tensors of perovskite, Li$_3$PO$_4$, and ZrO$_2$ datasets, respectively. Note that here we do not consider another popular mechanism of attention/transformers[62], although it may further improve the performance. Based on the size scaling, rather large systems become readily accessible, e.g., evaluation times of ~1 s could be attained for a structure with ~3000 atoms. This presents a prospect for studying a variety of local doping/substitution configurations quite efficiently, enabling the design of new materials.

We now reiterate some of the important aspects contributing to the high accuracy of the model, making it an efficient approach for "second-principles"[63] calculations of tensors of effective charge in oxides. The underlying approach is that of a convolution on a graph, whereby the target values can be gradually learned from the node and edge attributes by performing a sequence of interactions/convolutions with neighbor nodes. Message passing at each interaction allows the information to accumulate and propagate beyond nearest neighbors. Continuous filters are assigned to graph edges, enabling smooth modulation of trainable filter weights according to edge type. Equivariant edge features and descriptors allow continuous one-hot encoding and modulation of the convolution weights not only by the neighbor-neighbor distance, but also the direction of the interatomic vector, while equivariant node features naturally represent tensorial target values. The empirically established capability for regressing tensors of Born effective charges suggests sufficient expressiveness of the EGCNN architecture. In the future, an analysis of the components that are essential for performing efficient target tensor decomposition[64] should be performed.

Fundamentally, Born effective charges define the coupling of atomic vibrations to electric fields, and are responsible for long-range Coulomb interactions, which determine the TO-LO phonon mode splitting in polar crystals[44,65]. An early empirical model[66] yielded reasonable values for Born charges of various $ABO_3$ compounds from the fitting to experimental phonon mode oscillator strengths. With the advent of ab initio approaches, theoretical values have been calculated for many systems and found in good agreement with experiments based on comparison with infrared and inelastic neutron scattering spectra[67–70]. Recent developments include the extension of Born charges to metals[71,72], where they can be used to probe electron-phonon interactions.

Besides their utility for calculating physical observables, Born effective charges can also provide insights into the nature of microscopic interactions. Qualitatively, one can consider the two limits, one of purely electrostatic (or ionic), and the other of purely electronic interactions. In the former limit, a rigid ion model gives a good approximation of the interaction energy and effective charges. This ionic model of interactions is most applicable in $Li_3PO_4$, where the variation of the effective charges is smallest, indicating mostly ionic nature of bonding. This is especially true for Li, where the effective charge corresponds to the nominal value and shows small variability, whereas variability is somewhat larger in P and O.

On the other hand, perovskites and $ZrO_2$ display a mixture of covalent and ionic bonding. "Semicovalent" bonding is one of the characteristic features of perovskites, giving rise to several interesting phenomena, including the indirect (double) exchange coupling of cations[73]. The mixture of ionic and covalent bonding is responsible for anomalously high Born effective charges, as well as other disparities with the simple rigid ion model, such as the inequivalence of the charges of O anions[40,74]. This enhancement of Born effective charges has been traced to dynamical changes of O $2p$ and metal $d$ orbital hybridization[75] during ion displacement. The resulting flow of electrons augments the polarization change. Anomalous Born charges are also responsible for the large LO-TO gamma point optical phonon splitting[67] and appearance of the soft ferroelectric mode[76], and are central for achieving colossal permittivity materials[49,50]. We note that these effects are present in the studied systems and should be well captured by our EGCNN model.

## 6. Conclusion

In conclusion, we apply equivariant graph convolutional neural networks to learning the tensors of atomic Born effective charges. Good performance is achieved with three diverse systems: $ABO_3$ perovskites, $Li_3PO_4$, and $ZrO_2$. The model possibly demonstrates a good generalization ability by effectively learning from a small dataset, although a broader dataset[77] would provide a more rigorous test. Performance is further improved when using a larger dataset. The important hyperparameters affecting the model performance are the number of interactions/convolutions and the size of the descriptor in the interaction block. The training and inference times scale linearly with the number of atoms, and model architecture allows treating a wide range of combinations of chemical elements, enabling future creation of universal foundation models trained on a vast chemical space. Our model represents an important step towards fast and accurate modeling of microscopic quantum phenomena in response to the electric field.

Table I. Model performance summary.

|  | perovskites | $Li_3PO_4$ | $ZrO_2$ |
| --- | --- | --- | --- |
| base model BM1 (510k) MAE | 0.0207 | 0.0196 | 0.0323 |
| base model BM1 (510k) RMSE | 0.0300 | 0.0319 | 0.0510 |
| base model BM2 (140k) MAE | 0.0260 | 0.0227 | 0.0370 |
| base model BM2 (140k) RMSE | 0.0377 | 0.0364 | 0.0569 |
| specialized model (131k) MAE | 0.0088 | 0.0199 | 0.0325 |
| specialized model (131k) RMSE | 0.0135 | 0.0325 | 0.0493 |

**Data availability statement**
The datasets and weights of the pretrained models BM1 and BM2 supporting the findings of this study are openly available at the following URL/DOI: https://doi.org/10.17632/hx8kcpxh84.1 The python scripts for running evaluations with the model can also be downloaded from https://github.com/equivar/equivar_eval/

**Acknowledgement**
The work was supported by the JSPS Grant-in-Aid for Transformative Research Areas (A) (23H04105), and JST CREST "Nanomechanics" (JPMJCR1996). The computation was carried out using the general project on supercomputer "Flow" at Information Technology Center, Nagoya University.